\documentclass[aps,prl,reprint,groupedaddress]{revtex4-1}
\usepackage{graphicx}
\usepackage{ulem}

\begin{document}


\title{Delayed optical nonlinearity of thin metal films}


\author{Matteo Conforti}
\email[]{matteo.conforti@ing.unibs.it}
\affiliation{CNISM, Dipartimento di Ingegneria dell'Informazione, Universit\`a di Brescia, Via Branze 38, 25123 Brescia, Italy}
\author{Giuseppe Della Valle}
\affiliation{Dipartimento di Fisica and IFN-CNR, Politecnico di Milano, Piazza L. da Vinci 32, I-20133 Milan, Italy}

\date{\today}

\begin{abstract}
Metals typically have very large nonlinear susceptibilities,
whose origin is mainly of thermal character. We model the cubic nonlinearity of thin metal films by means of a delayed response derived  \textit{ab initio} from an improved version of the classic two temperature model. We validate our model by comparison with ultrafast pump-probe experiments on gold films.
\end{abstract}


\maketitle

Noble metal nanostructures, such as thin films, gratings, multilayers and nanoparticles, have been largely exploited in optics, especially within the fields of plasmonics and metamaterials. The linear optical response of these structures has been extensively studied both theoretically and experimentally, leading to the demonstration of unprecedented possibilities for extreme light concentration and manipulation (see \cite{Schuller10} and references therein). Also, the intense nonlinear optical response of metallic nano-structures after intense excitation with fs-laser pulses has attracted increasing attention in view of the potential to achieve ultra-fast all-optical control of light beams \cite{bennik99,yang04,husakou07,MacDonald08}.

Usually, the nonlinear response of the metal is modeled as a pure Kerr effect \cite{bennik99,yang04,husakou07}, where the nonlinear polarization is proportional to the cube of the electric field: $P_{NL}=\varepsilon_0\chi^{(3)}E^3$. In fact, this model though perfectly describing non resonant nonlinearities of electronic type that respond extremely fast (on the sub femtosecond time scale) to a driving electric field, turned out to be unsuitable for fs and ps optical pulses. Actually, the values of the $\chi^{(3)}$ coefficient, measured by the z-scan technique, that can be found in the literature, differ by up to two orders of magnitude~\cite{yang04,lee06,smith99, roten07}, clearly demonstrating that an instantaneous Kerr model is inadequate to describe the nonlinear response of metallic nanostructures. 

Pump-probe experiments in thin films \cite{sun94,groen95,hohlfeld00} and nanoparticles \cite{delfatti98,baida11} reveal that the nonlinearity of metals is due to the smearing of the electron distribution induced by intense optical absorption, resulting in a modulation of the inter-band and intra-band transition probabilities with subsequent variation of the dielectric permittivity. The temporal dynamics of the system, which has been accurately interpreted according to the two-temperature model (TTM) \cite{sun94,carpene06}, indicates that the nonlinear response is dominated by a delay mechanism, but a theoretical formulation in terms of a non-instantaneous $\chi^{(3)}$ polarization is still lacking.

In this Letter we derive, from an improved version of the TTM, a delayed third order non linear response suitable for the description of optically thin metallic structures. The outcome of our model is also quantitatively compared with experimental results from pump-probe spectroscopy on thin gold films.

Starting from Maxwell equations (written in MKS
units), neglecting transverse dimensions (i.e considering the
propagation of plane waves), we can obtain the 1D wave equation
for the electric field $E(z,t)$:
\begin{eqnarray}
\nonumber \frac{\partial^2 E(z,t)}{\partial z^2
}-\frac{1}{c^2}\frac{\partial^2}{\partial
t^2}\int_{-\infty}^{+\infty}E(z,t')\varepsilon(t-t')dt'=\\=\frac{1}{\varepsilon_0
c^2}\frac{\partial^2}{\partial t^2} P_{NL}(z,t),
\end{eqnarray}
\noindent where $c$ is the vacuum velocity of light, $\varepsilon_0$ is the
vacuum dielectric permittivity,
$\hat\varepsilon(\omega)=1+\hat\chi(\omega)$ and
$\hat\chi(\omega)$ is the linear electric susceptibility (\textit{hat} standing for Fourier transform).
In the perturbative regime, the nonlinear polarization can be expanded in Volterra series, accounting for small and non-istantaneous nonlinearity \cite{boyd}. Considering only third order nonlinearity, nonresonant, incoherent (intensity-dependent) nonlinear effects can be included by assuming the following functional form for the third-order polarization:
\begin{equation}
P_{NL}(z,t)=\varepsilon_0\left[\int_{-\infty}^{+\infty}\chi^{(3)}(t_1)E^2(t-t_1,z)dt_1\right] E(z,t)
\end{equation}

With the aim of quantifying the nonlinear response function by comparison with pump-probe experiments, we assume that the electric field is the sum of a powerful pump $A$ and a weak probe $B$: $E(z,t)=\frac{1}{2} A(z,t)e^{i\omega_at} +  \frac{1}{2} B(z,t)e^{i\omega_bt}+c.c.$, with $|A|>>|B|$.
In the slowly varying envelope approximation ($|\partial_t A|<<|A|$, $|\partial_t B|<<|B|$), the evolution equation for the probe becomes:
\begin{equation}\label{eq_probe}
\frac{\partial^2 B}{\partial z^2}+ k_b^2 B=-\frac{\omega^2_b}{2c^2}\int_{-\infty}^{+\infty}\chi^{(3)}(t-t_1)|A(t_1,z)|^2\,dt_1\,\,B
\end{equation}
where $k_b^2=\omega_b^2\varepsilon_b/c^2=\omega_b^2\hat\varepsilon(\omega_b)/c^2$. In optically thin metallic structures we can assume that the pump is non-depleted, so that the nonlinear response is space-independent. By doing so, Eq.~(\ref{eq_probe}) can be written as
\begin{equation}\label{wav_B}
\frac{\partial^2 B}{\partial z^2}+\frac{\omega_b^2}{c^2}\left[\varepsilon_b+\Delta\varepsilon (|A(t-t_d)|^2,\omega_b)\right]B=0,
\end{equation}
\noindent where the time-dependent nonlinear dielectric constant change $\Delta\varepsilon$ is a convolution between the pump pulse intensity and the third-order nonlinear response of the system ($t_d$ being the pump-probe delay). Equation (\ref{wav_B}) is a \textit{linear} wave equation for the probe $B$, where the time delay $t-t_d$ enters as a parameter. In the usual case of a short probe we can calculate the differential transmissivity (reflectivity) $\Delta T/T$ $(\Delta R/R)$ by considering $t=0$, where the probe is peaked.

The nonlinear response function $\chi^{(3)}(t)$ can be derived from the  extended two temperature model, that describes the evolution of the electrons and lattice temperature of a metal after absorption of a laser pulse \cite{sun94}:
\begin{eqnarray}\label{ttm}
\nonumber C_e\frac{\partial T_e}{\partial t}&=&-g(T_e-T_l)+a N\\
C_l\frac{\partial T_l}{\partial t}&=& g(T_e-T_l)+b N\\
\nonumber \frac{\partial N}{\partial t}&=&-aN-bN+P(z,t)
\end{eqnarray}
where $C_e$ and $C_l$ are the electronic and lattice heat capacities, $T_l$ is the lattice temperature, $g$ is the electron-phonon coupling constant, $N$ stands for the energy density stored in the nonthermalized part of the electronic distribution, $a$ is the electron gas heating rate and $b$ is the electron-phonon coupling rate. $P$ is the absorbed energy density and is related to field intensity $I$ through $P(z,t)=(1-R-T)\alpha e^{-\alpha z} I(t)$; for a thin film, we can neglect the spatial dependence and assume a mean absorbed energy density $P(t)=(1-R-T)I(t)/d$ ($d$, $R$, $T$, are the film thickness, reflection and transmission).
The energy density $N$ can be calculated as the convolution between the pump energy density and the thermalization response function $h_{th}(t)=\exp[-t/\tau_{th}]H(t)$ ($H(t)$ being the Heaviside function), with $\tau_{th}=1/(a+b)$.
The values $a,b$ can be derived from a more sophisticated extension of TTM proposed by Carpene \cite{carpene06}. It is worth noting that the two approaches give rather similar results, but the rate equation approach \cite{sun94} is simpler. We obtain $a=1/2\tau_1$, $\tau_1=E_F^2\tau_0/(\hbar\omega_b)^2$ and $b=1/\tau_{ep}$, where $E_F$ is the Fermi energy, $\tau_0=128\sqrt{3}\pi^2\omega_p$ ($\omega_p$ the plasma frequency), and $\tau_{ep}$ is the electron-phonon energy relaxation time. The value of $\tau_0$ has been calculated using the Lindhard dielectric function in the framework of Fermi liquid theory under the random phase approximation \cite{fann92}. This value turns out to be underestimated, mainly because of the $d$-band screening; a more correct value can be extracted from experimental data \cite{fann92}.
If $\Delta T_e= T_e-T_a<<T_a$, it is possible to consider the thermal capacity of electrons $C_e=\gamma T_e=\gamma T_0$ nearly constant: in this way the system (\ref{ttm}) is linear and can be solved exactly. $T_0$ can be estimated as a mean value between the minimum electronic temperature $T_a$ and the maximum value, that can be estimated by $T_{e,max}\approx\sqrt{T_a^2+2/\gamma\int P(t)dt}$.
For most metals, including Au and Ag,  $C_l>>C_e$, allowing us to write a simple expression for $\Delta T_e$:
\begin{equation}
\Delta T_e=\int N(t-t_1)\frac{C_e+a\tau_{th}C_le^{-t_1/\tau}}{C_e(C_e+C_l)\tau_{th}}H(t_1)dt_1
\end{equation}
i.e. the convolution between the energy density $N$ and the two temperature system response 
$h(t)=\frac{C_e+a\tau_{th}C_le^{-t/\tau}}{C_e(C_e+C_l)\tau_{th}}H(t)$.
It is now possible to calculate the total response of the extended two temperature system as the convolution of $h_{th}$ and $h$:
\begin{eqnarray}
&&h_{tot}(t)=\frac{1}{C_e+C_l}\times\\\nonumber&&\times\left[1-e^{-t/\tau_{th}}+\frac{aC_l\tau\tau_{th}}{C_e(\tau-\tau_{th})}\left(e^{-t/\tau}-e^{-t/\tau_{th}}\right)\right]H(t)
\end{eqnarray}
This way the electronic temperature change is simply calculated as the convolution between the pump and the total response: $\Delta T_e=\int P(t-t_1)h_{tot}(t_1)dt_1$. Assuming a gaussian pump $P(t)=P_0\exp[-2(t/T_0)^2]$ we obtain
\begin{widetext}
\begin{equation}\label{dTe}
\Delta T_e=\sqrt{\frac{\pi}{2}}\frac{P_0T_0/2}{C_e+C_l}\bigg\{
1+{\rm Erf} \left(\frac{\sqrt{2}t}{T_0}\right)+a\frac{C_l}{C_e}\frac{\tau\tau_{th}}{\tau-\tau_{th}}\bigg[e^{\frac{T_0^2}{8\tau^2}-\frac{t}{\tau}}{\rm Erfc}\left(\frac{\sqrt{2}}{T_0}\left(\frac{T_0^2}{4\tau}-t\right)\right)-
e^{\frac{T_0^2}{8\tau_{th}^2}-\frac{t}{\tau_{th}}}{\rm Erfc}\left( \frac{\sqrt{2}}{T_0}\left(\frac{T_0^2}{4\tau_{th}}-t\right)\right)\bigg]\bigg\}
\end{equation}
\end{widetext}
Following the same procedure a similar expression can be obtained also for the lattice temperature variation $\Delta T_L$  and the non-thermalized electron energy density $N$.

\begin{figure}
\includegraphics[width=7.5cm]{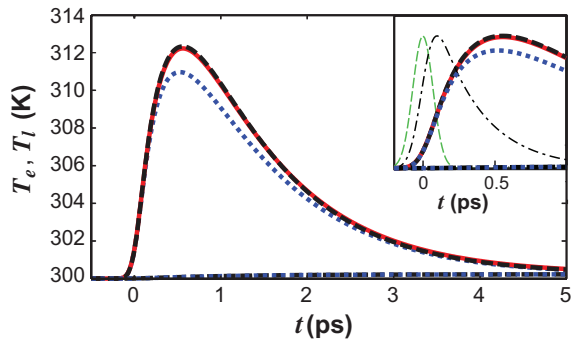}
\caption{ (color online) Temperature dynamics after irradiation of a thin gold film deposited on a sapphire substrate, from TTM of Eq.~(\ref{ttm}) (dashed black curves), from Carpene model \cite{carpene06} (dotted blue curves), and from analytical formula of Eq.~(\ref{dTe}) (solid red curve). Inset: zoom-in of the initial temperature dynamics; normalized energy density stored in non thermalized electrons (dash-dot black curve); pump intensity profile (thin dashed green curve). See text for parameters.}\label{fig1}
\end{figure}

We tested the validity of our approach by studying the temperature changes of a thin gold film ($d=20$~nm) deposited on a sapphire substrate ($n_s = 1.75$), after irradiation with a pump  pulse  of $\lambda_a=950$~nm wavelength, $T_0=140$~fs duration and $F=30$~$\mu$J/cm$^2$ fluence. Parameters for the TTM were taken from literature \cite{carpene06,hohlfeld00} as $T_a=300$~K, $\gamma=70$~J m$^{-3}$ K$^{-2}$, $C_l=2.5 \cdot 10^6$ J m$^{-3}$ K$^{-1}$, $g=2\cdot10^{16}$ W m$^{-3}$ K$^{-1}$, $\tau_{ep}=1.4$~ps, $\tau_0=6.5$~fs \cite{fann92}, $E_F=7.3$~eV. Figure \ref{fig1} shows the results of numerical solution of Eqs.~(\ref{ttm}), of numerical solution of Carpene model \cite{carpene06} and of analytical formula (\ref{dTe}). The agreement between analytical approximation and TTM is remarkable, indicating that in this range of electronic temperature the assumption of a constant heat capacity is reasonable. Carpene model and TTM give practically the same outcome for what concerns lattice temperature, but a small discrepancy can be noticed for what concerns the peak of electronic temperature. This feature is mainly due to the quite long thermalization time $\tau_{th}$ and will produce a slightly higher estimation of the maximum dielectric constant changes. In this figure we also show the time evolution of the energy density stored in the non thermalized electrons $N$. It acts like a delayed effective pump for electron and lattice, slowing the rise of $\Delta T_e$ and $\Delta T_l$. The dynamics of $N$ is faster than the thermalized one and it is responsible  for the small and quick dielectric permittivity changes observed when probing in a frequency range far from inter-band transitions \cite{sun94}.

Variation in electronic temperature induces variation in the dielectric constant through inter-band (bound electrons) transitions \cite{sun94}. 
Additional contributions to the change in the dielectric constant are given by intra-band transitions (that depend on lattice temperature) and by non thermalized electron energy distribution. These additional contributions are much smaller than the former and can be resolved only in spectral regions where inter-band transitions are not efficient \cite{sun94}. In order to obtain a reasonably simple model that however correctly grasps all the relevant phenomena, from now on we concentrate on inter-band transitions of thermalized electrons.

Energy injection into the conduction band smears the electron distribution around $E_F$ with reduction (increase) of the occupation probability of the electron states below (above) $E_F$. As a consequence, a modulation of inter-band transition probability is induced, with increased (decreased) absorption for transitions involving final states below (above) $E_F$. Inter-band transitions in noble metals are dominated by $d$-band to conduction band transitions near $L$ and $X$ points in the irreducible zone of the Brillouin cell. In the constant matrix element approximation, the variation of the imaginary part of the inter-band dielectric function of gold can be computed as follows \cite{Rosei_PRB74}:
\begin{eqnarray}\label{delta_epsi}
&&\Delta\varepsilon_2(\omega_b,T_e)=\frac{4 \pi^2 e^2}{3 m^2 \omega_b^2} \times \nonumber\\
&\times& \left[|P_L|^2 \Delta J_L(\omega_b,T_e) + |P_X|^2 \Delta J_X(\omega_b,T_e)\right]
\end{eqnarray}
\noindent where $m$ is the free electron mass, $P_L$ and $P_X$ are the electric-dipole matrix elements, and $\Delta J_L$ and $\Delta J_X$ the temperature induced variations of the \textit{Joint Density of States} (JDOS) for $d$-band to conduction band transitions near $L$ and $X$ respectively. Such variations can be computed as \cite{Rosei_PRB74}:
\begin{eqnarray}\label{delta_J}
\Delta J_{L,X}(\omega_b,T_e) &=& \nonumber \\
= \int_{E'_{L,X}}^{E''_{L,X}} &D_{L,X}&(E,\omega_b) \Delta P(E,T_e) dE
\end{eqnarray}
\noindent where $D_{L,X}(E,\omega_b)$ is the \textit{Energy Distribution of the Joint Density of States} (EDJDOS) of the considered transitions (with respect to the energy of final state $E$), and $\Delta P(E,T_e) = P(E,T_e) - P(E,T_a)$ is the temperature induced variation of the occupation probability for the final state (being $P(E,T_e) = 1-f(E,T_e)$ the probability that the final state is empty and $f(E,T_e)$ the Fermi-Dirac function). For small temperature changes  $\Delta T_e$, the following approximation holds:
\begin{eqnarray}\label{delta_P}
\Delta P(E,T_e) \simeq  \left[-\frac{\partial}{\partial T_e} f(E,T_e) \right]_{T_e = T_{a}} \Delta T_e.
\end{eqnarray}

\begin{figure}[hb!]
\includegraphics[width=7.2cm]{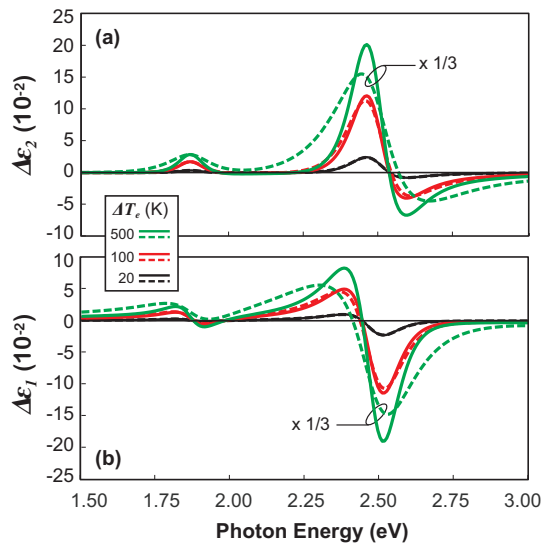}
\caption{ (color online) Spectral variation of the (a) imaginary part and (b) real part of the inter-band dielectric function of gold for three different values of $\Delta T_e$, computed from Eqs.~(\ref{delta_epsi}-\ref{delta_P}). Linearized temperature dependence provided by Eq.~(\ref{delta_P}) (solid lines) is compared with exact solution (dashed lines).}
\end{figure}

The EDJDOS was numerically computed under parabolic band approximation following the approach described in \cite{Rosei_PRB74}, taking effective masses, energy gaps, dipole matrix elements and integration limits $E'_{L,X}$ and $E''_{L,X}$ as reported in \cite{Guerrisi_PRB75}. The $\Delta \varepsilon_2$ resulting from Eqs.~(\ref{delta_epsi}-\ref{delta_P}) as a function of photon energy for three different temperature variations is shown in Fig.~2(a). The variation of the real part of the inter-band dielectric function $\Delta \varepsilon_1$ computed by Kramers-Kronig analysis of $\Delta \varepsilon_2$ is reported in Fig.~2(b). Note that the approximation provided by Eq.~(\ref{delta_P}) is accurate for $\Delta T_e$ as high as 100 K. 

For a quantitative validation of our model, we computed the transient differential reflection (transmission) $\Delta R / R$ ($\Delta T / T$) of the $20$-nm thin gold film according to standard thin film formulas~\cite{Abeles} and Eqs.~(\ref{dTe}-\ref{delta_P}). Results are reported in Fig.~3, compared with experimental data taken from the literature \cite{sun94} and full-numerical computation from the extended TTM. The probe photon energy is chosen at the maximum response of the thin film~\cite{sun94}, i.e. $\hbar \omega_b=2.48$~eV for $\Delta R / R$ and $\hbar \omega_b=2.43$~eV for $\Delta T / T$. We found that the analytical model is in quantitative agreement with experimental data, with only slight deviations in the long-time range ($t > 4$~ps) where the contribution from the lattice is dominant.

\begin{figure}[hb!]
\includegraphics[width=7.2cm]{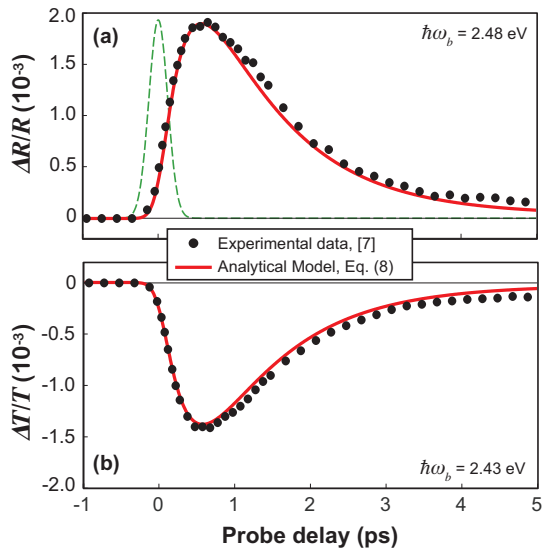}
\caption{ (color online) Transient (a) reflectivity and (b) transmissivity vs probe time delay, accounting for pump-probe cross-correlation (shown as dashed green line).}
\end{figure}

According to above analysis, a delayed cubic response of the metallic medium can be written as follows:
\begin{equation}
\chi^{(3)}(t)=\frac{\partial \Delta\varepsilon}{\partial T_e}\alpha c n \varepsilon_0h_{tot}(t)
\end{equation}
The magnitude of the nonlinear response can be estimated by assuming  continuous-wave pump and probe. In this case the nonlinear response is much faster than the fields, and the convolution reduces to the integral of $\chi^{(3)}(t)$. Assuming $C_l>>C_e$ we obtain:
\begin{equation}
\hat\chi^{(3)}(\omega_b;\omega_a,-\omega_a,\omega_b)=\frac{\partial \Delta\varepsilon(\omega_b)}{\partial T_e}\alpha(\omega_a) c n(\omega_a) \varepsilon_0\frac{a\tau_{th}\tau}{C_e}
\end{equation}
(the \textit{hat} is added because assuming c.w. fields is equivalent to give a frequency domain description of $\chi^{(3)}$). As an example, in the case of an infrared pump $\lambda_a=950$~nm and visible probe $\lambda_b=500$~nm, we obtain $\hat\chi^{(3)}(\omega_b;\omega_a,-\omega_a,\omega_b)\approx (-8.4 + 11i)\cdot10^{-8}$~esu~$=(-1.2 +1.5i)\cdot10^{-15}$~m$^2/$V$^2$. This huge value (six order of magnitudes greater than in fused silica) is due to the resonance of the probe with  interband transitions of gold. Therefore, it should be stressed that this kind of nonlinearity is strongly dispersive, and can be assumed constant only for reasonably band-limited pulses centered around $\omega_a$ and $\omega_b$. Changing either pump or probe frequency results in a drastic change of $\hat \chi^{(3)}(\omega_b)$, confirming that this kind of third order nonlinearity is not of Kerr-type. 

To conclude, we introduced a theoretical model for the delayed nonlinear response in optically thin noble-metal structures. A non-instantaneous $\chi^{(3)}$ coefficient is derived from an extended version of the TTM and semi-classical theory of optical transitions in solids. Our theoretical predictions turned out to be in quantitative agreement with experimental results from pump-probe experiments in thin gold films.

\end{document}